\documentclass[twocolumn]{emulateapj}
\usepackage{ amsmath, color, natbib,graphicx,tablefootnote,subfigure}
\usepackage{ stmaryrd }

\newcommand\ii{{\sc ii}}
\newcommand\iii{{\sc iii}}

\submitted{}
\shorttitle{CHAOS V: Carbon Abundances in M\,101}
\shortauthors{Skillman et al.}
\accepted{to ApJ}
\begin{document}  
\title{CHAOS V: Recombination Line Carbon Abundances in M\,101}
\author{
            Evan D.\ Skillman$^{1}$, 
            Danielle A.\ Berg$^{2,3}$, 
            Richard W.\ Pogge$^{2,3}$,  
            John Moustakas$^{4}$, 
            Noah S.\ J.\ Rogers$^{1}$, and 
            Kevin V.\ Croxall$^{2,5}$ 
}
\affil{
$^1$Minnesota Institute for Astrophysics, University of Minnesota, 116 Church St. SE, Minneapolis, MN 55455 \\
$^2$Department of Astronomy, The Ohio State University, 140 W 18th Ave., Columbus, OH, 43210 \\
$^3$Center for Cosmology \& AstroParticle Physics, The Ohio State University, 191 West Woodruff Avenue, Columbus, OH 43210 \\
$^4$Department of Physics \& Astronomy, Siena College, 515 Loudon Road, Loudonville, NY 12211\\
$^5$Illumination Works LLC, 5550 Blazar Parkway \#150, Dublin, OH 43017, USA 0000-0002-5258-7224\\
}

\begin{abstract}
The CHAOS project is building a large database of LBT \ion{H}{2} region spectra
in nearby spiral galaxies to use direct abundances to
better determine the dispersion in metallicity as a function of galactic radius.
Here, we present CHAOS LBT observations of \ion{C}{2}\,$\lambda$4267 emission detected in
10 \ion{H}{2} regions in M\,101, and, using a new photoionization model based
ionization correction factor, we convert these measurements into total carbon abundances.
A comparison with M\,101 \ion{C}{2} recombination line observations from the literature
shows excellent agreement, and we measure a relatively steep gradient in log(C/H)
of $-$0.37 $\pm$ 0.06 dex R$_e^{-1}$. 
The C/N observations are consistent with a constant value of log(C/N) $=$ 0.84 with a
dispersion of only 0.09 dex, which, given the different nucleosynthetic sources of C and N,
is challenging to understand.
We also note that when plotting N/O versus O/H, all of the \ion{H}{2} regions with
detections of \ion{C}{2}\,$\lambda$4267 present N/O abundances at the minimum of
the scatter in N/O at a given value of O/H.  
If the high surface brightness necessary for the detection
of the faint recombination lines is interpreted as an indicator of \ion{H}{2} region
youth, then this may point to a lack of nitrogen pollution in the youngest \ion{H}{2} regions.
In the future, we anticipate that the CHAOS project will significantly increase
the total number of \ion{C}{2}\,$\lambda$4267 measurements in extragalactic \ion{H}{2}
regions.

\end{abstract} 
 
 
 \section{Introduction}

 \subsection{The Nucleosynthetic Production of Carbon}

   The production of carbon is a relatively complex topic \citep[cf.][]{henry2000}.  
Although carbon is produced
primarily through the triple-$\alpha$ process, there are multiple sites where this process
can take place and the balance between carbon production and oxygen production has
dependencies on both nuclear physics and astrophysics \citep[cf.,][]{arnett1996}.  
Models of the chemical evolution in our galaxy indicate that massive stars and intermediate 
mass stars have played roughly equal roles in the production of carbon in
the solar vicinity \citep{carigi2005}.   

   Carbon clearly has at least two sites of production relevant to the chemical evolution of
galaxies.  The relatively flat relationship
between C/O and O/H observed at low metallicities 
\citep[e.g.,][]{garnett1995, garnett1997, izotov1999, esteban2014, berg2016, berg2019} 
is indicative of a primary production
process in massive stars \citep[e.g.,][]{woosley1995}.
   The observed increase in the C/O ratio with increasing metallicity 
\citep[e.g.,][]{garnett1999, esteban2005, esteban2009, esteban2019} indicates that secondary-like
production becomes important at higher metallicities.  Although this is similar to the 
observed trend of increasing N/O with increasing metallicity, that relationship is 
understood as a direct consequence of the CNO cycle.  There is no similar nucleosynthetic
process for the production of C, so the observed relative constancy of C/N 
is suggestive of a physical conspiracy of sorts. The increase of C/O with increasing O/H has
been attributed to both an increasing yield of C in massive stars with increasing
metallicity due to strengthening of stellar winds \citep[e.g.,][]{maeder1992, meynet2002, hirschi2005},
and metallicity dependent yields of intermediate mass stars \citep[e.g.,][]{vandenhooek1997}.

  Carbon abundances in unevolved Galactic stars have the promise of tracing the evolution of 
both absolute and relative abundances as a function of time and place within our galaxy
\citep[e.g.,][]{nissen2014}, and carbon abundances
in evolved stars can trace the nucleosynthetic processes involved with the later
stages of stellar evolution \citep[e.g.,][]{henry2018, feuillet2018, hasselquist2019}.  
Alternatively, carbon abundances from the ISM provide
only the cumulative endpoint of all nucleosynthesis, gas flows, and mixing, 
but their observation can still
give constraints on the chemical evolution of galaxies. Hopefully these constraints can lead to a 
better understanding of the dominant carbon production processes.  Here we will
present ISM carbon abundances derived from carbon recombination lines.  

\begin{deluxetable*}{lccccc}
\tabletypesize{\scriptsize}
\tablecaption{M\,101 \ion{H}{2} Regions with CHAOS \ion{C}{2}\,$\lambda$4267 Detections}
\tablewidth{0pt}
\tablehead{
  \colhead{$\frac{\rm{R}}{\rm{R}_{25}}$}        &
  \colhead{\ion{H}{2} Region Name}        &
  \colhead{CHAOS ID\tablenotemark{a}}        &
  \colhead{I(C\,\ii\,$\lambda$4267)/I(H$\beta$)}        &
  \colhead{T[O\,\iii] (K)\tablenotemark{a}}     &
  \colhead{T[S\,\iii] (K)\tablenotemark{a}}
  }
\startdata
0.196 & H1013\tablenotemark{b}     &    +164.6+009.9     &  0.0023 $\pm$  0.0003 &  7420 $\pm$   180  &     6910 $\pm$   110  \\
0.284 & H1052\tablenotemark{b}     &    +189.2$-$136.3   &  0.0029 $\pm$  0.0002 &  7769 $\pm$    77  &     8900 $\pm$   160  \\
0.335 & NGC 5461  &    +254.6$-$107.2   &  0.0013 $\pm$  0.0004 &  8739 $\pm$    66  &     9340 $\pm$   320  \\
0.426 & NGC 5462  &    +354.1+071.2     &  0.0009 $\pm$  0.0003 &  9542 $\pm$    90  &     8190 $\pm$   260  \\
0.435 & NGC 5462  &    +360.9+075.3     &  0.0013 $\pm$  0.0003 &  9179 $\pm$    93  &     8030 $\pm$   280   \\
0.468 & NGC 5455  &    $-$099.6$-$388.0 &  0.0010 $\pm$  0.0002 &  9443 $\pm$    70  &     8920 $\pm$   210  \\
0.541 & NGC 5447  &    $-$368.3$-$285.6 &  0.0011 $\pm$  0.0002 &  9299 $\pm$    69  &    11150 $\pm$   270  \\
0.554 & NGC 5447  &    $-$392.0$-$270.1 &  0.0012 $\pm$  0.0004 &  9579 $\pm$    75  &    11200 $\pm$   290  \\
0.669 & H1216\tablenotemark{b}     &    +509.5+264.1     &  0.0004 $\pm$  0.0004 & 10692 $\pm$    88  &     9930 $\pm$   240  \\
0.813 & NGC 5471  &    +667.9+174.1     &  0.0004 $\pm$  0.0003 & 12790 $\pm$   170  &    12110 $\pm$   410
\enddata
\label{table1}
\tablenotetext{a}{from \citet{croxall2016}}
\tablenotetext{b}{naming convention from \citet{hodge1990}}
\end{deluxetable*}

 \subsection{The Abundance Discrepancy Factor (ADF)}

  Because there are no strong collisionally excited lines from carbon in the optical
regime, carbon abundance determinations in \ion{H}{2} regions are very limited.  The two most popular
alternatives are the collisionally excited semi-forbidden lines of CIII] at $\lambda\lambda$1907, 1909 
in the ultraviolet \citep[e.g.,][]{garnett1995, berg2016, berg2019} 
and the \ion{C}{2} recombination line at $\lambda$4267
\citep[e.g.,][]{esteban1998, esteban2002, esteban2005, peimbert2005, 
bresolin2007, esteban2009, esteban2014, toribio2016, toribio2017, esteban2019}.  
The primary observational challenge of observing the collisionally excited ultraviolet lines, 
beyond needing space-based 
observations, is the lack of a nearby hydrogen emission line to use as a reference.  
The solution to this is to compare to the nearby OIII] lines at $\lambda\lambda$ 1660, 1666.  
The primary observational challenge for the optical recombination lines is the intrinsic extreme 
weakness of the lines which are typically observed at $\sim$0.1\% of H$\beta$ (with a roughly
linear dependence on O/H).

  However, with sufficient sensitivity to observe the C recombination line, a second
difficulty arises.  In principle, absolute ionic abundances from recombination lines
should be very robust because of the negligible dependence on temperature when converting 
from line flux ratios to abundances.  However, whenever recombination line
ionic abundances are compared to ionic abundances derived from collisionally excited lines
(most frequently for O$^{++}$ and C$^{++}$) an ``abundance discrepancy factor''
(ADF) with an amplitude of roughly a factor of two is found \citep[in the sense that the 
recombination lines produce higher abundances, cf.][]{garcia2007}.  Given the inherent bias to higher 
values in the determination of temperatures from auroral-to-nebular flux ratios
of collisionally excited lines (the basis of the ``direct'' abundance method) and
the high temperature sensitivity of the relative strengths of the collisionally
excited lines, the most natural explanation for the ADF is that there are
temperature fluctuations in the \ion{H}{2} regions \citep [e.g.,][]{peimbert1967, peimbert1969}.  
Note that this explanation is currently  debated in the literature; 
the most common criticism is that the size of the 
temperature fluctuations required to explain the observed ADF is larger than 
can easily be reproduced in photoionization models \citep[e.g.,][]{simondiaz2011,stasinska2013}.   

  In this paper, we will derive ionic C$^{++}$/H$^{+}$ abundances from the \ion{C}{2}\,$\lambda$4267 
line and then convert the ionic abundance to a total abundance, C/H, via ionization corrections
based on photoionization modeling. We also reference our derived C/H values to previously 
reported O/H and N/H values derived from collisionally excited lines
using the direct method.  Because the C$^{++}$/H$^{+}$ ADF is of order a factor of two \citep[e.g.,][]{toribio2017},
our C/O and C/N ratios are expected to be offset upward by 
roughly a similar factor of two when comparing them to ratios determined purely from  
recombination lines or collisionally excited lines.

\begin{deluxetable*}{llccccccc}
\tabletypesize{\scriptsize}
\tablecaption{M\,101 CHAOS \ion{H}{2} Region Carbon Abundances}
\tablewidth{0pt}
\tablehead{
  \colhead{$\frac{\rm{R}}{\rm{R}_{25}}$}        &
  \colhead{Region Name} &
  \colhead{\rm{C$^{++}$}/\rm{H$^{+}$}}  &
  \colhead{\rm{O$^{++}$}/\rm{O}\tablenotemark{a}}  &
  \colhead{C/H ICF}             &
  \colhead{12+log(\rm{C}/\rm{H})}  &
  \colhead{log(\rm{C}/\rm{O})}  &
  \colhead{log(\rm{C}/\rm{N})}  &
  \colhead{12+log(\rm{O}/\rm{H})\tablenotemark{a}} \\
 & & \colhead{$\times$ 10$^{-5}$} & & & & & &
  }
\startdata
0.196 &  H1013    & 18.3 $\pm$  2.2 & 0.30 $\pm$ 0.02 & 0.54 $\pm$  0.05 & 8.53 $\pm$ 0.06 & -0.04 $\pm$  0.07 &  0.92 $\pm$  0.08 &  8.57 $\pm$  0.02 \\
0.284 &  H1052    & 23.7 $\pm$  1.5 & 0.66 $\pm$ 0.02 & 0.76 $\pm$  0.05 & 8.49 $\pm$ 0.04 & -0.08 $\pm$  0.04 &  0.93 $\pm$  0.05 &  8.57 $\pm$  0.01 \\
0.335 &  NGC 5461 & 11.2 $\pm$  2.9 & 0.66 $\pm$ 0.03 & 0.76 $\pm$  0.05 & 8.17 $\pm$ 0.12 & -0.31 $\pm$  0.12 &  0.69 $\pm$  0.13 &  8.48 $\pm$  0.02 \\
0.426 &  NGC 5462 &  7.6 $\pm$  2.9 & 0.51 $\pm$ 0.06 & 0.66 $\pm$  0.06 & 8.06 $\pm$ 0.18 & -0.39 $\pm$  0.19 &  0.81 $\pm$  0.21 &  8.45 $\pm$  0.05 \\
0.435 &  NGC 5462 & 10.5 $\pm$  2.7 & 0.64 $\pm$ 0.02 & 0.74 $\pm$  0.05 & 8.15 $\pm$ 0.12 & -0.28 $\pm$  0.12 &  0.90 $\pm$  0.13 &  8.43 $\pm$  0.01 \\
0.468 &  NGC 5455 &  8.7 $\pm$  2.0 & 0.70 $\pm$ 0.03 & 0.79 $\pm$  0.05 & 8.04 $\pm$ 0.11 & -0.34 $\pm$  0.11 &  0.81 $\pm$  0.13 &  8.39 $\pm$  0.02 \\
0.541 &  NGC 5447 &  9.1 $\pm$  1.8 & 0.56 $\pm$ 0.02 & 0.69 $\pm$  0.05 & 8.12 $\pm$ 0.09 & -0.30 $\pm$  0.09 &  0.83 $\pm$  0.10 &  8.42 $\pm$  0.01 \\
0.554 &  NGC 5447 &  9.8 $\pm$  3.6 & 0.69 $\pm$ 0.02 & 0.78 $\pm$  0.05 & 8.10 $\pm$ 0.17 & -0.25 $\pm$  0.17 &  0.88 $\pm$  0.18 &  8.35 $\pm$  0.01 \\
0.669 &  H1216    &  3.3 $\pm$  3.3 & 0.65 $\pm$ 0.05 & 0.75 $\pm$  0.06 & 7.64 $\pm$ 0.50 & -0.63 $\pm$  0.50 &  0.70 $\pm$  0.50 &  8.26 $\pm$  0.03  \\
0.813 &  NGC 5471 &  4.0 $\pm$  3.1 & 0.78 $\pm$ 0.05 & 0.85 $\pm$  0.05 & 7.67 $\pm$ 0.46 & -0.47 $\pm$  0.46 &  0.90 $\pm$  0.50 &  8.14 $\pm$  0.03
\enddata
\label{table2}
\tablenotetext{a}{from \citet{croxall2016}}
\end{deluxetable*}

  \subsection{CHAOS Observations of M\,101}

While numerous spectra of star forming galaxies have been obtained through large surveys 
such as SDSS \citep{strauss2002,tremonti2004}, PINGS \citep{pings}, CALIFA \citep{sanchez2012}, 
MaNGA \citep{bundy2015,law2015}, SAMI \citep{bryant2015}, TYPHOON \citep[e.g.,][]{ho2017},
and PHANGS-MUSE \citep{kreckel2019}, few of these observations 
enable direct determinations of absolute gas-phase abundances, as they do not detect 
the faint auroral lines that reveal the electron temperatures of the \ion{H}{2} regions.  
Even determining the relative abundances can be challenging given possible biases, 
both in the observations and the methodology of determining gas-phase metallicity 
using only the brightest lines \citep{kewley2008}.  Furthermore, the coarse spatial 
resolution that results from observing distant galaxies means that non-homogeneous 
clouds of gas will co-inhabit each spectrum \citep{moustakas2006}.  The goal of the
CHemical Abundances Of Spirals (CHAOS) program is to obtain very high quality spectra 
of \ion{H}{2} regions in nearby spiral galaxies, which allow direct determinations of absolute 
and relative abundances across a broad range of parameter space \citep[e.g.,][]{berg2015}.
The CHAOS program uses the Multi-Object Double Spectrographs \citep[MODS,][]{mods} 
on the Large Binocular Telescope \citep[LBT,][]{hill2010} to observe large numbers of \ion{H}{2} regions
in spiral galaxies.  
Recognizing that direct abundances do have systematic biases \cite[e.g.,][]{peimbert1967,stasinska2005},
our goal is to obtain a very large sample of direct abundances as a best chance to 
assess those systematics.

\begin{figure}[tbp]
\epsscale{1.0}
   \centering
   \plotone{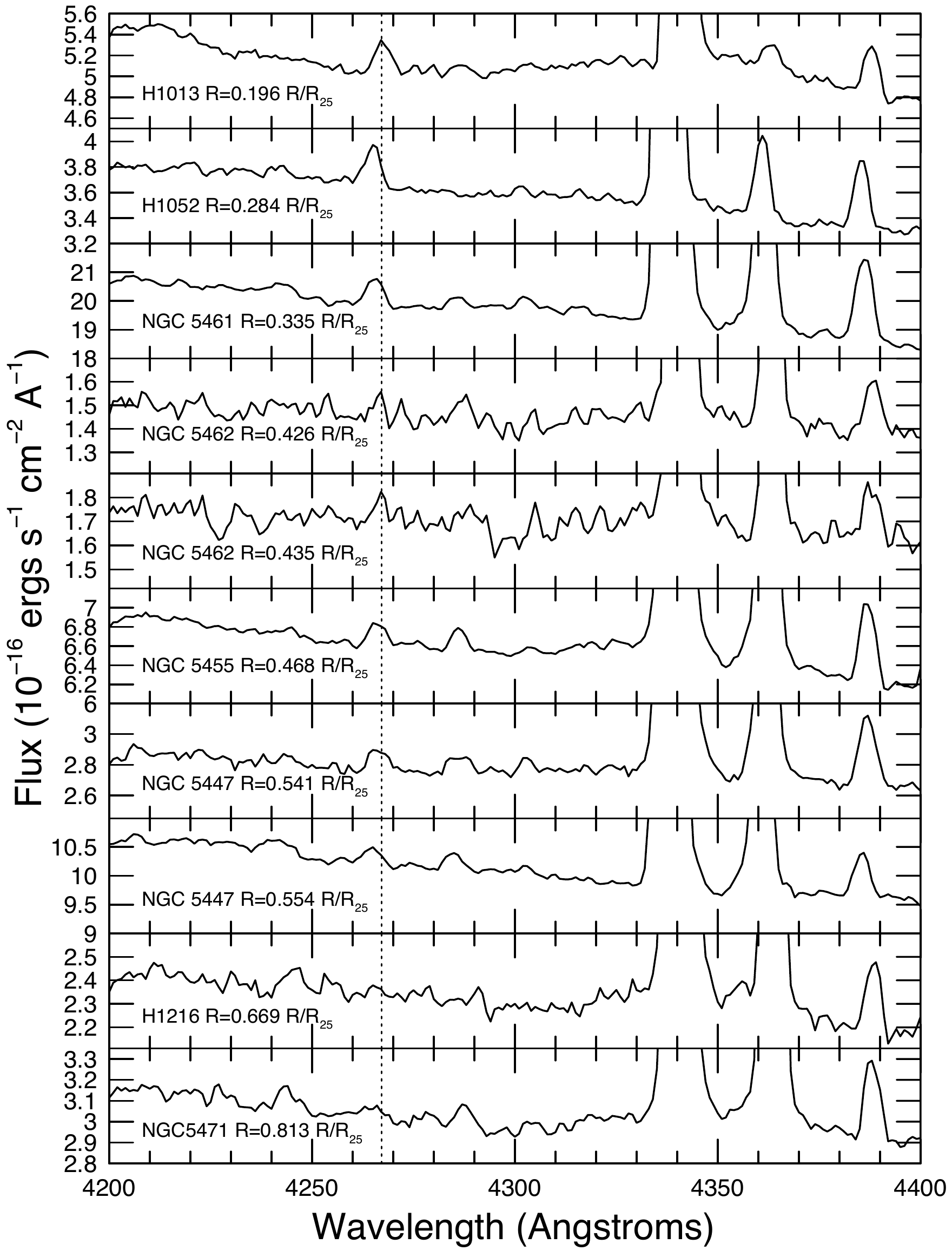}
   \caption{CHAOS spectra of the M\,101 \ion{H}{2} regions, where the \ion{C}{2}\,$\lambda$4267
recombination line was detected, zoomed in
to the region surrounding that line. The $\lambda$4267 line is marked with a vertical dotted line.
The three strong lines at longer wavelengths are \ion{H}{1}\,$\lambda$4340, [\ion{O}{3}]\,$\lambda$4363,
and \ion{He}{1}\,$\lambda$4388.  The spectra are arranged in increasing galactocentric radius from
top to bottom.  Note that, at this scaling, the \ion{He}{1}\,$\lambda$4388 line is relatively constant
while the \ion{C}{2}\,$\lambda$4267 line becomes weaker with increasing radius (decreasing metallicity)
and the [\ion{O}{3}]\,$\lambda$4363 line becomes stronger with increasing radius.
}
   \label{fig:spectra}
\end{figure}

The nearly face-on spiral galaxy M\,101 (NGC\,5457), with its relatively steep metallicity gradient, 
is an ideal target for \ion{H}{2} region observations and has been observed several times 
\citep[e.g.,][]{kennicutt2003, bresolin2007, li2013, esteban2019}.
As part of CHAOS, \citet{croxall2016} presented observations of 74 \ion{H}{2} regions in M\,101 
with direct abundances.  Although it was not an original goal of the CHAOS project, we 
realized that our 
observations were sensitive enough to detect the \ion{C}{2}\,$\lambda$4267 recombination line
in a number of our spectra.  Here we present an analysis of those observations.

Our observations and data reduction are described in~\S2.  
In~\S3 we determine gas-phase chemical abundances.  
We present abundance gradients for C/H, C/O, and C/N in~\S4.  
We discuss relative abundances and the origin of C in~\S5.   
Finally, we summarize our conclusions in~\S6.
We will assume the same properties for M\,101 as in \citet{croxall2016}, i.e., 
a distance of 7.4 Mpc \citep{ferrarese2000} (with a resulting scale of 35.9 pc/arcsec), 
R$_{25}$ $=$ 864\arcsec \citep{kingfish},
an inclination angle of 18\degr , and a major-axis position angle of 39\degr \citep{things}.
We also use a value for R$_e$ of 198\arcsec \citep{berg2020}. 

\section {CHAOS Spectra and Reductions}

The observations presented here were previously reported in \cite{croxall2016} and we provide a brief summary.
Optical spectra of M\,101 were taken using MODS on the LBT during the spring semester of 2015.  
All spectra were acquired with the MODS1 unit.  
We obtained simultaneous blue and red spectra using the G400L (400 lines mm$^{-1}$, 
R$\approx$1850) and G670L (250 lines mm$^{-1}$, R$\approx$2300) gratings, respectively.  
The main goal of this program was to detect the intrinsically weak auroral lines in the 
wavelength range from 3200\,--\,10,000 \AA\  
in order to obtain direct abundances.  Although not a design goal of the program, 
the observations were sensitive enough
to detect recombination line emission from the \ion{C}{2} recombination line at $\lambda$4267.  
Recombination line observations are normally made at higher spectral resolution than in the 
CHAOS program, but because the \ion{C}{2}\,$\lambda$4267 is 
well isolated, these detections are possible. 
 
Figure \ref{fig:spectra} shows the wavelength region of the spectra highlighting 
the \ion{C}{2}\,$\lambda$4267 emission 
line for the 10 \ion{H}{2} regions in M\,101 where it was detected.  Previously, in the CHAOS program, 
we have limited reporting of 
detections to a minimum signal/noise cutoff of 3.  Because of the extreme faintness of the 
\ion{C}{2}\,$\lambda$4267 emission line and the very high worth of any detections, we have dropped that 
constraint \citep[as is customary in the recombination line literature, e.g.,][]{esteban2009} for the
two lowest metallicity detections.
However, instead of marking reported weak detections with colons, we have reported our best estimates
of the uncertainties of the measurements. Note also that in two cases, two of our observations
are independent regions in large \ion{H}{2} region complexes (i.e., NGC~5462 and NGC~5447).  

We also searched the spectra for the presence of
\ion{O}{2} multiplet 1 recombination lines (e.g., $\lambda\lambda$ 4639, 4642, 4649, 4651, 4662, 
4674, 4676), but here our
relatively low spectral resolution of $\sim$ 2 \AA\ limited our sensitivity to these 
lines.  Not only were some of these
lines confused with emission from \ion{N}{3} at $\lambda$4641 and [\ion{Fe}{3}] at $\lambda$4658,
but broad Wolf-Rayet ``blue bump'' emission between $\lambda$ 4630 and 4700 \AA\ was detected in all
of the spectra in which \ion{C}{2}\,$\lambda$4267 emission was detected.  As a result, we cannot report
any C/O measurements based solely on recombination lines.

Detailed descriptions of the data reduction procedures are provided in \citet{berg2015} and \citet{croxall2016}.  
The measurements of the \ion{C}{2}\,$\lambda$4267 emission lines were made identically to the measurements reported in
\citet{croxall2016}.  The \ion{C}{2}\,$\lambda$4267 line was not measured as part of the original CHAOS pipeline because we 
did not anticipate very many detections.
For the analysis here, we use the de-reddened relative emission line strengths reported in 
\citet{croxall2016} and the physical conditions (e.g., temperatures and densities) reported therein.
Given the weakness of the line, the associated uncertainty is dominated by photon counting statistics.
In Table~\ref{table1}, we report locations of the \ion{H}{2} regions where \ion{C}{2}\,$\lambda$4267 has been
detected, the reddening corrected \ion{C}{2}\,$\lambda$4267 emission line fluxes relative to H$\beta$, and the 
[\ion{O}{3}] and [\ion{S}{3}] temperatures from \citet{croxall2016}.  

\begin{figure*}[tp]
\epsscale{1.}
   \centering
   \plotone{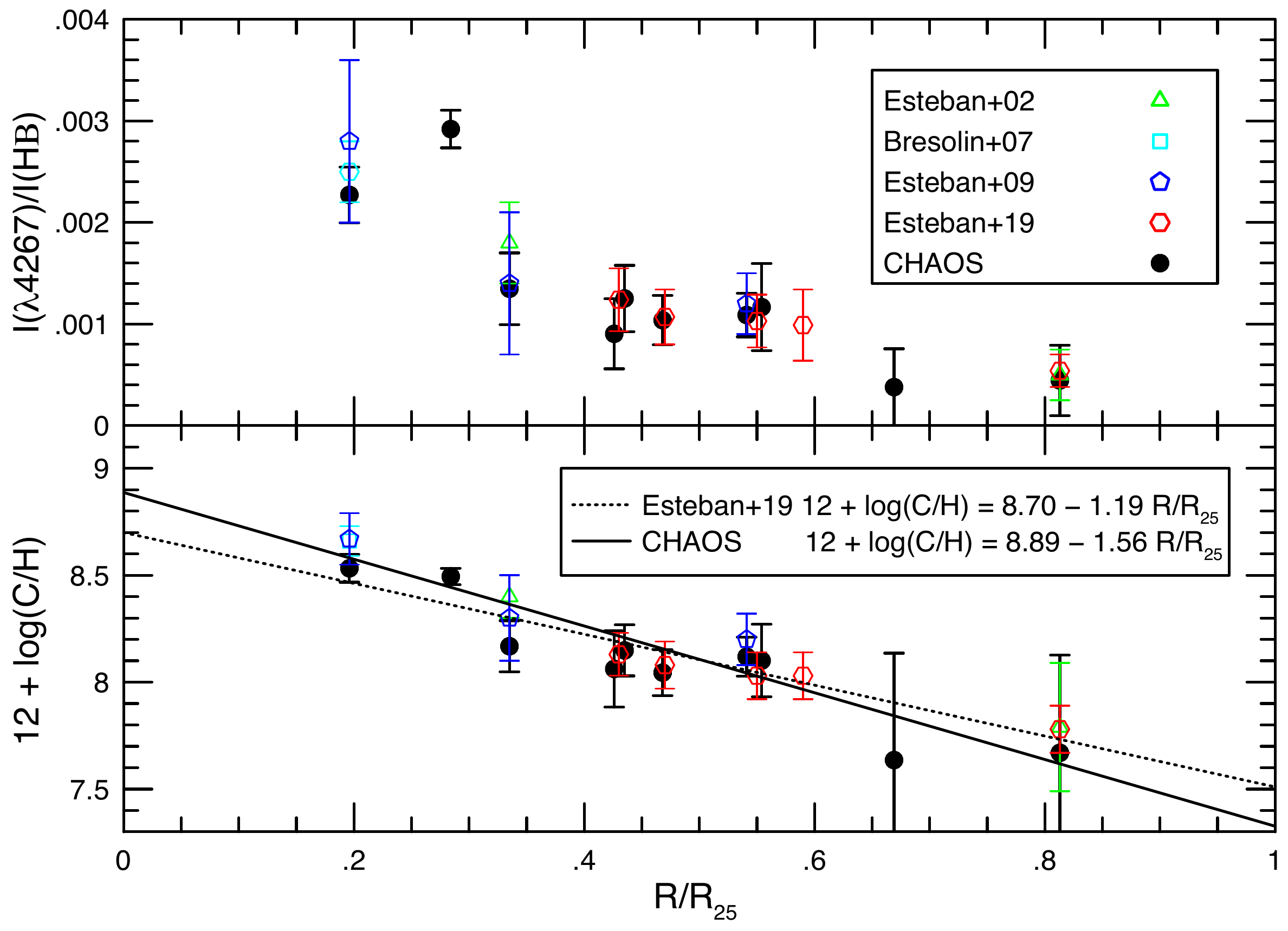}
   \caption{The top panel shows our \ion{C}{2}\,$\lambda$4267/H$\beta$ observations in 10 M\,101 \ion{H}{2} 
regions plotted
as a function of radius.  We have included previously published observations \citep[from][]{esteban2002, 
bresolin2007, esteban2009, esteban2019} for comparison. The agreement with previous observations is very good.  
In the bottom panel we plot the corresponding values of log(C/H) as a function of radius.  The bottom panel also shows 
a comparison to the values of log(C/H) from the literature.  
The solid line shows a fit to the CHAOS data and the dotted line shows the fit from \citet{esteban2019}. 
 }
  \label{fig:ionic}
\end{figure*}

In the top panel of Figure~\ref{fig:ionic}, we plot our \ion{C}{2}\,$\lambda$4267 emission
line fluxes relative to H$\beta$ as a function of the deprojected distances from the galaxy center in 
units of the isophotal radius, R$_{25}$ \citep{croxall2016}.
Because of the
weak dependence on temperature, the $\lambda$4267/H$\beta$ ratio directly reflects the
C$^{++}$/H$^{+}$ ratio and one immediately sees evidence of the strong radial gradient.  We have
added to this plot the other published observations of \ion{C}{2}\,$\lambda$4267 emission in M\,101 for
comparison \citep{esteban2002, bresolin2007, esteban2009, esteban2019}. 
The top panel of Figure~\ref{fig:ionic} shows the
excellent agreement of our observations with the ten previously published detections
in six distinct \ion{H}{2} regions (for consistency, we have used the radial distances adopted
in this work). In sum, we have added four \ion{C}{2}\,$\lambda$4267 observations of \ion{H}{2} regions
lacking prior detections and \ion{C}{2}\,$\lambda$4267 observations for two of these regions were 
subsequently reported by \citet{esteban2019}.

\section{Gas-Phase Carbon Abundances}

\subsection{Ionic Abundances}

The $\lambda$4267/H$\beta$ ratios can be converted directly to C$^{++}$/H$^{+}$ ratios with knowledge
of their respective emissivities.  \citet{davey2000} have calculated recombination coefficients for
\ion{C}{2} lines, and in their Table 5, they provide polynomial fits at an electron density of
10$^4$  cm$^{-3}$ over the temperature range of 5,000K to 20,000K.  We use their Case B formula for making
our calculations.   \citet{hummer} have calculated the corresponding recombination coefficients
for \ion{H}{1}.  Although the \ion{H}{2} regions in M\,101 are observed to have lower densities than
the 10$^4$ cm$^{-3}$ used in the emissivity calculations, the effective recombination coefficient
is not very sensitive to density at these temperatures \citep{davey2000}.
In our calculations, we 
do not attribute uncertainties to the atomic data as the observational uncertainties are
considered to be much larger.  Although assessing uncertainties on atomic data from calculations
can be difficult, our assumption is based mostly on the report by \citet{davey2000} that a
comparison of the effective recombination coefficient for the \ion{C}{2}\,$\lambda$4267 line with
the earlier calculation by \citet{pequignot1991} agrees within 2.4\%.  Note also, that if the 
cause of the ADF is nebular temperature fluctuations, and if a significant amount of the 
recombination line emission is produced
in significantly cooler regions than measured by the auroral to nebular line ratios,
then the temperature range of 5,000K to 20,000K of the calculations by \citet{davey2000} may be inadequate. 

Although the temperature dependence of the conversion from flux ratio to ionic abundance is 
quite small (roughly a 10\% variation over the temperature range of the \ion{H}{2} regions), 
a nebular temperature is required to make the calculation.  The natural temperature to use would
be that derived from the [\ion{O}{3}] $\lambda$4363/$\lambda$5007 ratio. 
Since there has been some speculation regarding the reliability of nebular temperatures 
derived from the [\ion{O}{3}] $\lambda$4363/$\lambda$5007 ratio \citep[e.g.,][]{binette2012, berg2015, berg2020},
we report in Table~\ref{table1} the T[\ion{O}{3}] and T[\ion{S}{3}] values for each of the 
\ion{H}{2} regions calculated by \citet{croxall2016}.  From Table~\ref{table1} it can be seen that,
although the measurements differ by many times their statistical uncertainties,  
there is coarse agreement between the two measurements.  
Most importantly, given the relatively large observational uncertainties on the recombination lines
and the very low dependence of the abundances on the electron temperature, 
the choice of temperature will not have any impact.  
For simplicity, we have chosen to use T[\ion{O}{3}] for this calculation.
The values of C$^{++}$/H$^{+}$ and their uncertainties are reported in Table~\ref{table2}.
 
\begin{figure*}[tp]
\epsscale{1.}
   \centering
   \plotone{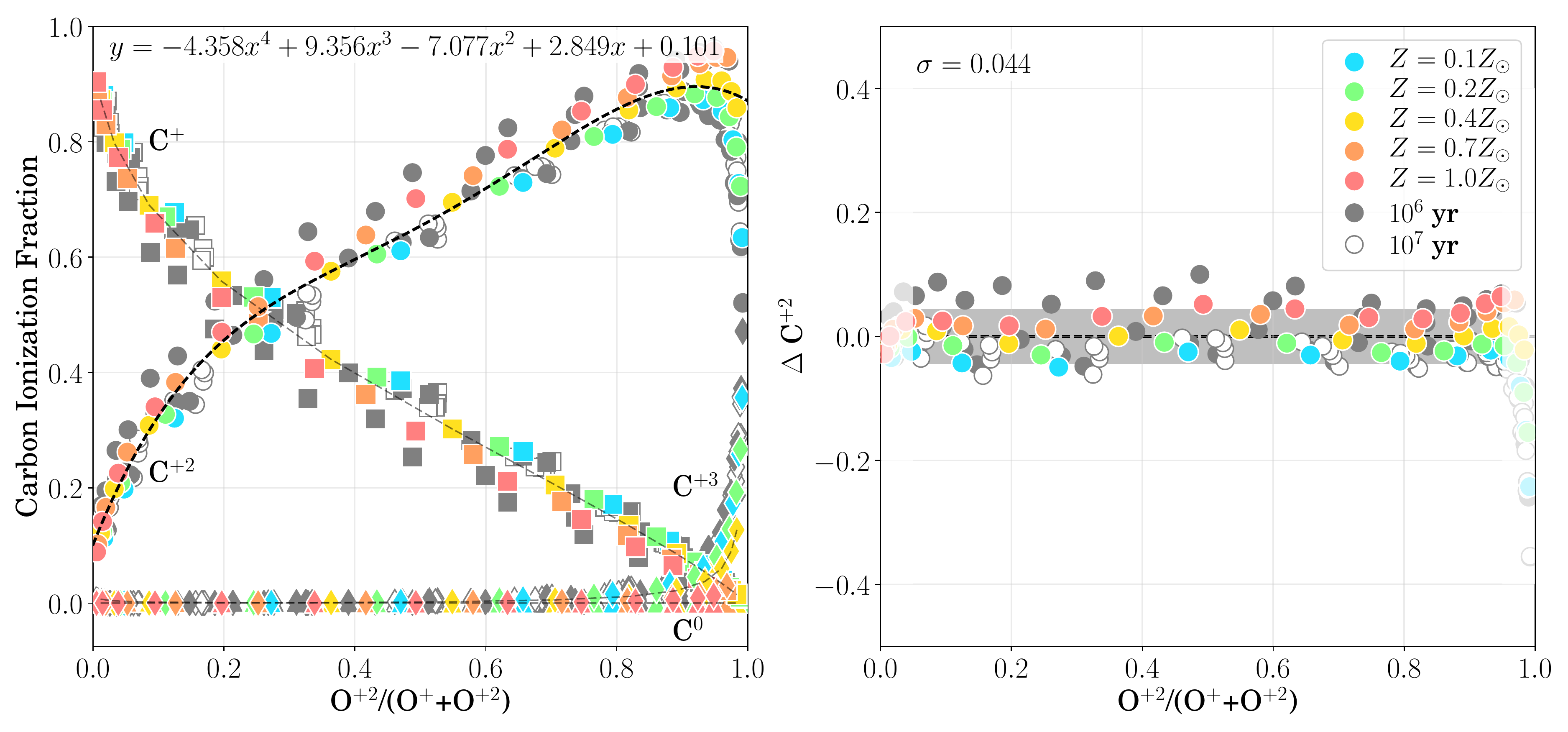}
   \caption{Left panel: C species ionization fractions as a function of O$^{++}$ ionization fraction for 
the purpose of calculating C ionization correction factors. 
Grey symbols designate deviations due to burst age ranging from from t = 10$^{6}$ (filled symbols) to t = 10$^{7}$ (open symbols) years. 
Solid symbols are color-coded by the gas-phase oxygen abundance at t = 10$^{6.7}$ yr. 
The different ionic species are designated by triangles, squares, circles, and diamonds in order of increasing ionization. 
Dashed gray lines connect the Z = 0.4 × Z$_{\odot}$ models, or 12 + log(O/H) = 8.3, which is a typical nebular 
abundance for our sample. 
Right panel: The dispersion in C$^{++}$ ionization fraction as a function of O$^{++}$ ionization fraction for
all of the models presented in the left panel.  The total dispersion is 0.044 from 0.05 $\le$ O$^{++}$/O $\le$ 0.95.
 }
  \label{fig:icf}
\end{figure*}

\subsection{Ionization Correction Factors and Total Abundances}

Ionization correction factors (ICFs) are required to convert our C$^{++}$/H$^+$ measurements
to values of C/H.  Because of the small variations in the ionization correction 
for converting C$^{++}$/O$^{++}$ to C/O \citep{garnett1995, berg2016, berg2019, esteban2019}, and the relative 
robustness of nebular O/H measurements, one possible choice would be to compare our C$^{++}$/H$^+$ measurements
to O$^{++}$/H$^+$ measurements and to then convert C$^{++}$/O$^{++}$ to C/O, which can then be  
multiplied by O/H to obtain the C/H ratio.  This is the method typically used when observing both 
C$^{++}$ and O$^{++}$ in either optical recombination lines or ultraviolet collisionally excited lines.
This would be possible here, although the resultant C/O would come from a mix of recombination lines
and collisionally excited lines (because we have no O$^{++}$ recombination line observations).

Alternatively, we can use the ionization corrections derived from photoionization models to 
go directly from C$^{++}$/H$^+$ to C/H.  
Figure~\ref{fig:icf} shows an expanded grid of models from that reported in \citet{berg2019}
which was investigated to judge the stability of the corresponding ICF.   We found that 
this ICF is very well behaved over the
oxygen fractional ionization interval 0.05 $<$ O$^{+2}$/(O$^{+}$ + O$^{+2}$) $<$ 0.95, 
over a large range of metallicities and \ion{H}{2} region ages, 
with a dispersion in the models of only 4.4\%.  
In this range,
the ICF varies smoothly from $\sim$0.3 to $\sim$ 0.9. While this range in ICF is larger than the the range 
of ICFs based on 
C$^{++}$/O$^{++}$, the strong stability of the ICF over this range in oxygen fractional ionization
gives us confidence that this is a satisfactory choice.  From this photoionization modeling,
we derive the following relationship:
\begin{equation}
C^{+2}/H^+ = 0.101 + 2.849 x -7.077 x^2 +9.356 x^3 -4.358 x^4
\end{equation}
where x is the oxygen fractional ionization $=$ O$^{+2}$/(O$^{+}$ + O$^{+2}$).  The ICFs are reported in Table~\ref{table2} 
along with the corresponding values of log(C/H).   The uncertainties on the ICFs represent both 
the observational uncertainties on the values of the ionization fractions of oxygen (which are required 
to calculate the ICFs) and the dispersion in photoionization models, added in quadrature.   

With values of C/H, we can then multiply by our previously calculated
values of O/H and N/H to obtain C/N and C/O, which are also tabulated in Table~\ref{table2}.  
Note that because the O/H and N/H values are calculated from collisionally
excited lines and the C/H values from recombination lines, that the resulting C/O and C/N values are
roughly 0.3 dex higher than values typically reported in the literature \citep[cf.][]{esteban2019}.
We are also not making any corrections 
for dust. Although both carbon and oxygen are known to be depleted onto dust grains, typically
the corrections for both are of order 0.1 dex.  For example, in the comprehensive study by \citet{esteban1998}, 
the upward gas abundance to gas-plus-dust abundance corrections for the Orion nebula are 0.10 dex for carbon
and 0.08 dex for oxygen.  In the study of absorption in Milky Way lines of sight by
\citet{jenkins2004} the relative ease with which an individual element depletes are nearly identical
for C and O (in Jenkins' formalism, $-$0.097 versus $-$0.089), although the offset from these relations 
is $\sim$\,0.1 dex higher for C than for O ($-$0.148 versus $-$0.050). 
So, very generally, the gas-phase C/O ratio is probably indicative of the total C/O ratio, but there
is a fair degree of uncertainty in this assumption.  

\section{Abundance Gradients}
\subsection{The C/H Abundance Gradient}

A minimum number of \ion{H}{2} regions spanning a significant fraction of the disk are
required to produce an accurate measurement of an abundance gradient 
\citep[e.g.,][]{zaritsky1994, skillman1996, rosolowsky2008, bresolin2009, moustakas2010}.
With 10 \ion{H}{2} regions spanning from 0.2 to 0.8 in R/R$_{25}$, we have sufficient
observations to significantly constrain abundance gradients.  However, at
the largest radii, the uncertainties on the very weak detections of the
 \ion{C}{2}\,$\lambda$4267 line do mitigate the constraints on the abundance gradients.
Note that the recent measurement of C/H in NGC~5471 by \citet{esteban2019} has 
a vastly improved uncertainty. 
Additional measurements of the ultraviolet \ion{C}{3}] in the outer M\,101 regions
would be welcome in this regard (although the ADF would need to be accounted for). 
 
In the bottom panel of Figure~\ref{fig:ionic} we have plotted log(C/H) as a function of 
galactic radius.  From a weighted fit to the CHAOS data we find:
\begin{equation} 12 + \rm{log(C/H) = }
        \begin{array}{l}8.89 (\pm0.09) - 1.56 (\pm0.25)~\rm{(R/R}_{25})\\
                               8.89 (\pm0.09) - 0.050(\pm0.008)~\rm{(R/kpc)}\\
                               8.89 (\pm0.09) - 0.37 (\pm0.06)~\rm{(R/R_e)}
        \end{array},
\end{equation} 
and we have plotted this relationship as a solid line in Figure~\ref{fig:ionic}. 

We have also plotted the previously reported values of log(C/H) from 
the literature \citep{esteban2002, bresolin2007, esteban2009, esteban2019}.  As should follow from the 
agreement with the observed fluxes, the derived values of log(C/H)
are in good agreement.  Note that for the innermost \ion{H}{2} region, 
H1013, where the CHAOS value of 8.53 $\pm$ 0.06 is lower that the 8.67 $\pm$ 0.12 previously
reported by \citet{esteban2009}, that \citet{esteban2019} have calculated a new value of
8.58 $\pm$ 0.11 by applying a new ICF, resulting in agreement.  
Similarly, for NGC~5461, the CHAOS value of 8.17 $\pm$ 0.12
is in agreement with the value of 8.21 $\pm$ 0.16 which was revised down from the 
original report of 8.30 $\pm$ 0.20 in \citet{esteban2009}.
Overall, this indicates that our CHAOS measurements, although obtained at a lower
spectral resolution, are capable of delivering quality C/H abundances. 

\citet{esteban2009} derived a linear best fit to the radial gradient of C/H in M\,101 of
12 + log (C/H) = 8.90 $-$1.32 ($\pm$ 0.33) R/R$_{25}$, in agreement 
with our value. \cite{esteban2019} find 
12 + log (C/H) = 8.70 ($\pm$ 0.10) $-$ 1.19 ($\pm$ 0.19) R/R$_{25}$, with the lower
intercept and slope, relative to the earlier result of \citet{esteban2009}, due, 
in part, to the lowering of the C/H abundances in 
their inner two \ion{H}{2} regions and the very small uncertainty in their
new measurement of NGC~5471.  We have added the C/H radial gradient from
\citet{esteban2019} to Figure~\ref{fig:ionic} as a dotted line.  As expected 
from the agreement between the measurements, the gradients are in 
agreement.  For all future comparisons to literature data in this paper,
we will compare solely to the \citet{esteban2019} results.

\begin{figure}[tbp]
\epsscale{1.0}
   \centering
   \plotone{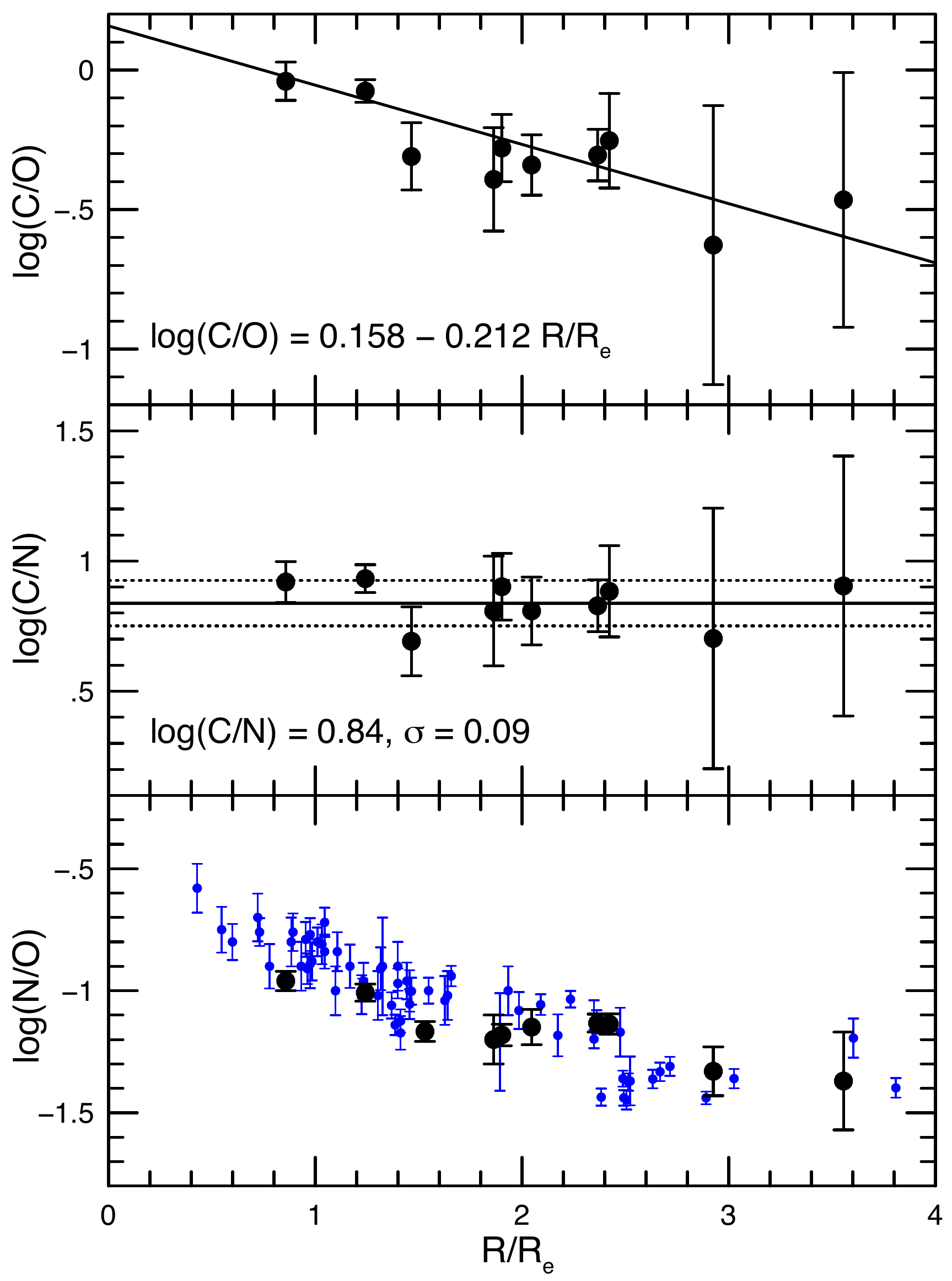}
   \caption{
Radial abundance gradients of C/O, C/N, and N/O in M\,101.  The top two panels show the results
from our \ion{C}{2}\,$\lambda$4267 observations.  The bottom plot shows the N/O values presented in
\citet{croxall2016} with the black color coding indicating where we have detected \ion{C}{2}\,$\lambda$4267
and all other points indicated in blue. The bottom plot shows 
that the regions detected in \ion{C}{2}\,$\lambda$4267 are representative of the whole sample.} 
  \label{fig:grad}
\end{figure} 

\subsection{Relative Abundance Gradients}

In Figure~\ref{fig:grad}, 
we plot the radial gradients for the C/O, C/N, and N/O abundance ratios from the CHAOS \ion{H}{2} 
regions in M\,101.
In Figure~\ref{fig:grad}, we have switched from using R/R$_{25}$ to R/R$_e$ on the abscissa because of
the evidence for more universal behavior of radial abundance gradients when comparing 
galaxy gradients based on the effective radius. 
Indeed, \citet{sanchez2014} and \citet{sanchez2016} have proposed a 
characteristic oxygen abundance gradient
in spiral galaxies when measured in terms of the half-light or effective radius
\citep[but see also][]{sanchez2018}, and \citet{berg2020} have shown evidence for
similar slopes in N/O.

In the top two panels of Figure~\ref{fig:grad}, we plot the data made possible 
by the new measurements.  The bottom panel reproduces the values from \citet{croxall2016}
for all of the M\,101 \ion{H}{2} regions with direct abundances.
Comparing the radial distribution coverage in 
the top two panels with that in the bottom panel demonstrates that we have been able to sample
most of the observable disk of M\,101 with our \ion{C}{2}\,$\lambda$4267 observations.

In the top panel of Figure~\ref{fig:grad}, we see that log(C/O) has a significant negative
gradient and a weighted linear fit yields
\begin{equation} 12 + \rm{log(C/O) = }
        \begin{array}{l}0.158 (\pm0.087) - 0.212 (\pm0.057)~\rm{(R/R}_{25})\\
                               0.158 (\pm0.087) - 0.0068 (\pm0.0018)~\rm{(R/kpc)}\\
                               0.158 (\pm0.087) - 0.048 (\pm0.013)~\rm{(R/R_e)}
        \end{array}.
\end{equation} 
This log(C/O) gradient is different from 
the value of log(C/O) $=$ $-$0.16 ($\pm$ 0.24) $-$ 0.40 ($\pm$ 0.46) R/R$_{25}$ 
found by \citet{esteban2019}, but that is to be expected because our O/H values are
based on observations of collisionally excited lines. The intercept is higher because
of the expected $\sim$ 0.3 dex offset between the direct and the recombination line O/H values.
The smaller slope reflects a metallicity dependence in these values, although the slopes
are consistent within the uncertainties. 

In contrast,
the middle panel of Figure~\ref{fig:grad} shows that log(C/N) is consistent with a constant 
value.  The unweighted mean value of log(C/N) is 0.84 with a dispersion of 0.09 dex 
(the weighted mean is nearly identical at 0.86).
This relatively constant value of C/N over a significant range in metallicity presents
a challenge to explain given the different nucleosynthetic sources of C and N.  

The relative
constancy of log(C/N) means that log(C/O) and log(N/O) should have similar dependencies.
This is not immediately obvious from a comparison of the top and bottom panels of Figure~\ref{fig:grad}.
The data for log(N/O), shown in the bottom panel of Figure~\ref{fig:grad},
are characterized by a decline with increasing radius followed by a 
relative flattening at large radii, while the data for log(C/O) appear to be well described by a linear fit.
However, considering the lack of C/H measurements in the inner 0.20 R/R$_{25}$ and 
the small number of outer points (two) with relatively large uncertainties on the measurements of log(C/O),
the data are certainly consistent with a flattening in log(C/O) at larger radii.  The new measurement in
NGC~5471 by \citet{esteban2019} is interesting in this regard because of the smaller uncertainty which 
does, indeed, favor a flattening the the C/O radial gradient in M\,101.

\begin{figure}[tbp]
\epsscale{1.0}
   \centering
   \plotone{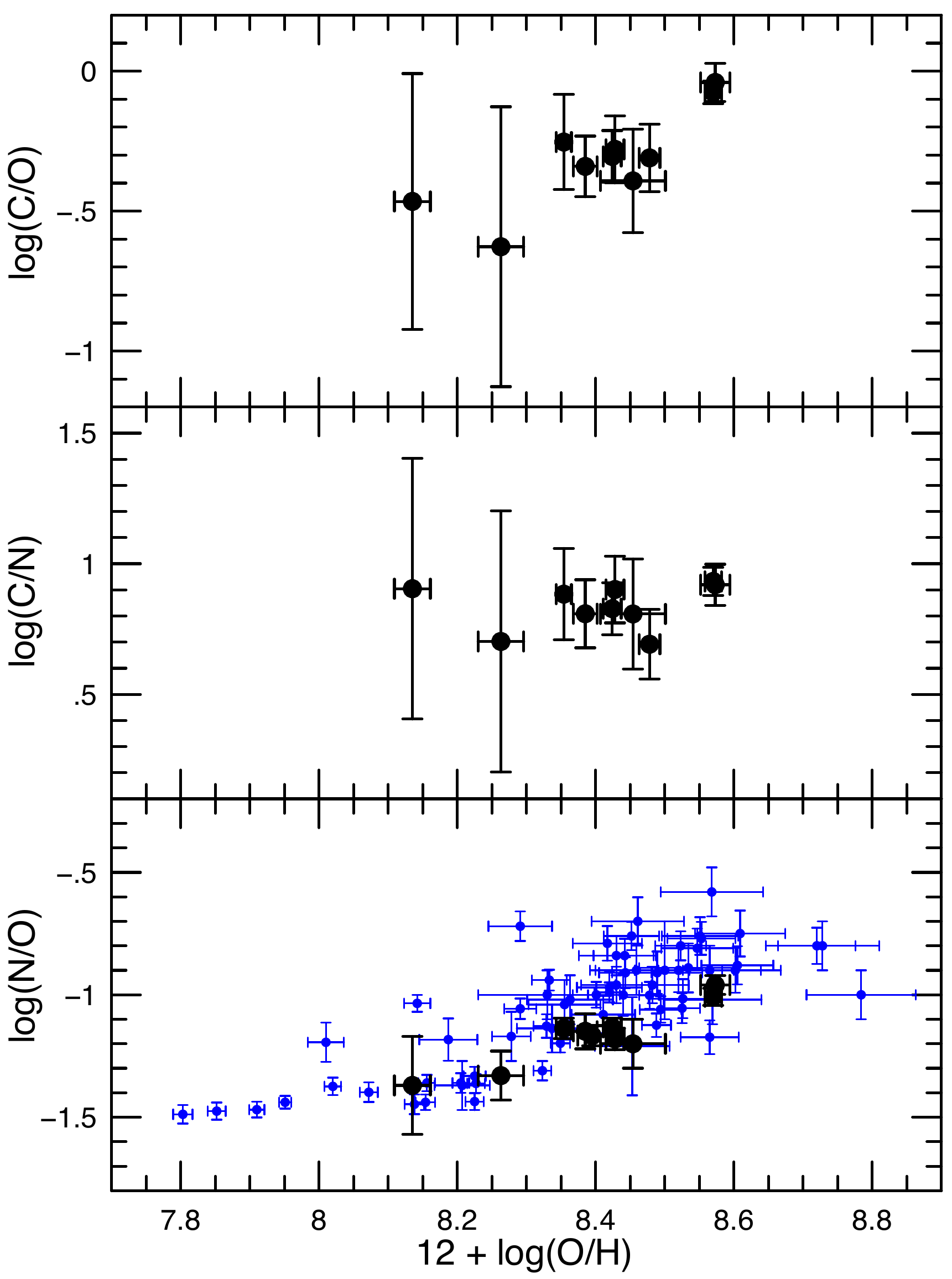}
   \caption{
Abundances of C/O, C/N, and N/O plotted versus O/H in M\,101.  
As in Figure~\ref{fig:grad} the top two panels show the results
from our \ion{C}{2}\,$\lambda$4267 observations and the bottom plot shows the N/O values presented in
\citet{croxall2016}.  In the bottom plot, the points where we have detected \ion{C}{2}\,$\lambda$4267 are indicated
in black and the other points from \citet{croxall2016} are indicated in blue. The bottom plot shows
that the regions detected in \ion{C}{2}\,$\lambda$4267 cover a somewhat limited range in O/H and are located
preferentially at lower values of N/O for a given O/H.
}
   \label{fig:rel}
\end{figure}

\section{Relative Abundance Trends}

\subsection{Trends with metallicity}

In Figure~\ref{fig:rel}, we re-plot the data from Figure~\ref{fig:grad} reordered by log(O/H) 
instead of by galactocentric radius.  As in Figure~\ref{fig:grad}, the top two panels show the results
from our new \ion{C}{2}\,$\lambda$4267 observations and the bottom plot shows the N/O values presented in
\citet{croxall2016}.  The top two panels of Figure~\ref{fig:rel} reproduce the trends seen in Figure~\ref{fig:grad}, with 
a significant slope in log(C/O) and relatively constant values of log(C/N).  

However, the bottom plot in Figure~\ref{fig:rel} reveals something potentially very interesting.
As in Figure~\ref{fig:grad}, the points where we have detected \ion{C}{2}\,$\lambda$4267 are indicated
in black and the other points from \citet{croxall2016} are indicated in blue. The bottom plot shows
that the regions detected in \ion{C}{2}\,$\lambda$4267 cover a somewhat limited range in O/H, i.e., although
the last measured point in radius is very near the radial limit of the observations, there are
several \ion{H}{2} regions with lower values of O/H extending roughly 0.3 dex lower than the 
last measured point with a \ion{C}{2}\,$\lambda$4267 detection.  This is likely simply due to the bias
of weakening \ion{C}{2}\,$\lambda$4267 emission at lower values of O/H. An interesting impression from
Figure~\ref{fig:rel} is the distribution of \ion{C}{2}\,$\lambda$4267 detection points relative to the 
rest of the \ion{H}{2} regions.  Clearly N/O shows a large scatter,
but the \ion{C}{2}\,$\lambda$4267 detection points all appear to be located
preferentially at lower values of N/O for a given O/H.  These points tend to delineate the 
lower bound in N/O in the N/O versus O/H plot.  Given that the \ion{C}{2}\,$\lambda$4267 detections 
represent the highest surface brightness (and typically the youngest) \ion{H}{2} regions,
this might either represent an observational bias or something more physical like 
a lack of nitrogen pollution in the youngest objects.

In a comparison of CHAOS observations of four spiral galaxies, \citet{berg2020} found a strong trend
such that the \ion{H}{2} regions with the lowest N/O ratios at a given value of O/H had the highest values 
of the O$^{+2}$ fractional ionization, O$^{+2}$/(O$^{+}$ + O$^{+2}$).   If the high values of
the O$^{+2}$ fractional ionization are indicative of the youngest ionizing stellar clusters, 
then the higher values
of N/O in \ion{H}{2} regions with lower O$^{+2}$ fractional ionization (and thus, indicative of
older age stellar ionizing clusters) might be simply understood in terms of localized nitrogen
pollution \citep[e.g.][]{kobulnicky1997}.  
Note that this trend is opposite to the original suggestion of \citet{garnett1990} that low values of N/O
are indicative of recent pollution by supernovae which are then followed by higher values of N/O
when the nitrogen production of intermediate mass stars has had a chance to catch up.  

\subsection{On the Origins of Carbon}

Insights into the nucleosynthesis of the elements is often provided by trends in relative
and/or absolute abundances.
The relatively constant value of C/N over a large range in metallicity remains a
confounding challenge to chemical evolution theories which require a fine tuning of the
contributions of C from massive and intermediate mass stars to match the nucleosynthesis
of N which comes predominantly from the CNO process at the metallicities typical of
spiral galaxies.

\citet{esteban2019} have gathered the literature data for five spiral galaxies with 
nebular carbon abundance measurements and find a variety of C/H abundance gradients 
as a function of R$_e$.  In the future, the CHAOS project has the potential to double the
number of galaxies with measured C/H abundance gradients and perhaps comparisons of
the C, N, and O abundances in these galaxies with different chemical evolutionary histories
will provide better insight into the nucleosynthetic origins of carbon.

\section {Conclusions}

CHAOS project LBT/MODS observations of \ion{H}{2} regions in M\,101 have resulted in
detections of the \ion{C}{2}\,$\lambda$4267 recombination line in ten regions.
These observations are in excellent agreement with the previously published 
observations for six regions in M\,101.

We use a newly calculated ionization correction factor for C$^{++}$/H$^{+}$ to convert
our C$^{++}$ abundances to total C abundances.

These new observations allow us to provide secure measurements of the C/H and C/O
gradients in M\,101.  The C/H gradient in M\,101 is relatively steep with a slope in 
log(C/H) of $-$0.37 $\pm$ 0.06 dex R$_e$$^{-1}$, in good agreement with the previous
measurements by \citet{esteban2009, esteban2019}.
The C/N observations are consistent with a constant value of log(C/N) $=$ 0.84 with a
dispersion of only 0.09 dex.  The constant C/N ratio over a large range in metallicity (0.5 dex)
is challenging to understand, given the different nucleosynthetic sources of C and N.  

When plotted as a function of O/H, the trends in C/O and C/N are quite similar to 
what is seen in the radial gradients.  That C/N is relatively constant implies that
C/O and N/O should be similar. The large number of N/O observations show a flattening 
at large radii while the current C/O observations are insufficient to significantly
constrain the trend for the outer \ion{H}{2} regions.
Additional observations of C in the outer \ion{H}{2}
regions of M\,101 will be very useful in this regard.

We also note that when plotting N/O versus O/H, all of the \ion{H}{2} regions with
detections of \ion{C}{2}\,$\lambda$4267 present N/O abundances at the minimum of
the scatter in N/O at a given value of O/H.  If this is not the result of an 
observational bias, and if the high surface brightness necessary for the detection 
of the faint recombination lines is interpreted as an indicator of \ion{H}{2} region
youth, then this may point to a lack of nitrogen pollution in the youngest \ion{H}{2} regions.

Due to the extreme weakness of the carbon recombination lines, our contribution represents a 
valuable addition to the study of carbon in the ISM of spiral galaxies.  In the future, we
will analyze and publish additional \ion{C}{2} emission line detections from other galaxies in
the CHAOS project.

\acknowledgments
This work has been supported by NSF Grants AST-1109066 and AST-1714204.  
E.D.S. is extremely grateful for the opportunity to attend the Second Workshop 
``Chemical Abundances in Gaseous Nebulae: Open Problems in Nebular Astrophysics'' held at
the Universidade do Vale do Paraiba, Brazil, in March 2019 where he learned 
many things important to the development of this paper.
We thank the anonymous referee for a very prompt and insightful report which we believe 
significantly improved this paper.
This paper uses data taken with the MODS spectrographs built with funding from NSF grant AST-9987045 and the NSF Telescope System Instrumentation Program (TSIP), with additional funds from the Ohio Board of Regents and the Ohio State University Office of Research.  
This paper made use of the modsIDL spectral data reduction pipeline developed in part with funds provided by NSF Grant AST-1108693.  
We are grateful to D. Fanning, J.\,X. Prochaska, J. Hennawi, C. Markwardt, and M. Williams, and others who have developed the IDL libraries of which we have made use: coyote graphics, XIDL, idlutils, MPFIT, and impro.  
This work was based in part on observations made with the Large Binocular Telescope (LBT). The LBT is an international collaboration among institutions in the United States, Italy and Germany. The LBT Corporation partners are: the University of Arizona on behalf of the Arizona university system; the Istituto Nazionale di Astrofisica, Italy; the LBT Beteiligungsgesellschaft, Germany, representing the Max Planck Society, the Astrophysical Institute Potsdam, and Heidelberg University; the Ohio State University; and the Research Corporation, on behalf of the University of Notre Dame, the University of Minnesota, and the University of Virginia.
This research has made use of the NASA/IPAC Extragalactic Database (NED) which is operated by the 
Jet Propulsion Laboratory, California Institute of Technology, under contract with the 
National Aeronautics and Space Administration and the NASA Astrophysics Data System (ADS).

\end{document}